\documentclass[12pt]{article}
\usepackage{cite,graphicx,amsmath,amssymb}
\usepackage{graphicx}
\usepackage{multicol}
\usepackage{xcolor}
\usepackage{slashed}
\usepackage[normalem]{ulem}
\usepackage{bbm}
\usepackage{hyperref}
\usepackage{cancel}

\newcommand{\be}{\begin{equation}}
\newcommand{\ee}{\end{equation}}
\newcommand{\bea}{\begin{eqnarray}}
\newcommand{\eea}{\end{eqnarray}}
\newcommand{\al}[1]{\begin{align}#1\end{align}}

\newcommand{\paren}[1]{\left(#1\right)}

\newcommand{\ol}{\overline}

\newcommand{\nn}{\nonumber\\}

\def\g{\gamma}

\def\p{\partial}

\def\ol{\overline}

\def\*{\dagger}
\def\I{\mathbbm{1}}
\def\K{\mathcal{K}}

\begin{document}

\thispagestyle{empty}
\begin{flushright}
CTPU-PTC-21-23, MAD-TH-21-01
\end{flushright}

\vfil
\vspace*{-.2cm}

\begin{center}

{\Large\bf Completing the D7-brane local gaugino action}
\\[2cm]

{\large Yuta Hamada$^{1}$, Arthur Hebecker$^{2}$, Gary Shiu$^{3}$ and Pablo Soler$^{4}$}

\vspace{1cm}

{\it

${}^{1}$ {Department of Physics, Harvard University, Cambridge, MA 02138 USA}\\[.1cm]
 ${}^{2}$ Institute for Theoretical Physics, University of Heidelberg, 
Philosophenweg 19,\\ D-69120 Heidelberg, Germany\\[.1cm]
${}^{3}$ Department of Physics, University of Wisconsin-Madison, Madison, WI 53706, USA\\[.1cm]
${}^{4}$ Center for Theoretical Physics of the Universe, Institute for Basic Science,\\ Daejeon 34051, South Korea
}

\vspace{.3cm}
\large{May 24, 2021}
\\[1.6cm]

{\bf Abstract}
\end{center} 

Within the ongoing debate about de Sitter (dS) vacua in string theory, different aspects of explicit dS proposals have come under intense scrutiny. One key ingredient is D7-brane gaugino condensation, which is usually treated using effective 4d supergravity. However, it is clearly more desirable to derive the relevant scalar potential directly from a local 10d Lagrangian. Such a local 10d description captures the interactions among the various localized sources and the background fields which are smeared in the 4d Lagrangian. While progress in this endeavour has recently been made, some form of non-locality related to the 4-gaugino term has remained hidden in the available proposals. We spell out the local counterterm removing the divergence that arises when integrating out the 3-form flux and which, upon dimensional reduction, serves to reproduce the relevant part of the 4d supergravity action.  This is both a step towards a more complete understanding of 10d type-IIB supergravity as well as specifically towards better control of dS constructions in string theory involving gaugino condensation.

\newpage

\tableofcontents 

\vspace*{1cm}

\section{Introduction}

Non-perturbative effects play a decisive role in string phenomenology. 
They have long been used to stabilize moduli in string theory (see e.g.~\cite{Derendinger:1985kk, Dine:1985rz} and Ch.~15 of\cite{Ibanez:2012zz} for further refs.). In more recent times, gaugino condensation on D7-branes has been invoked as a key ingredient in the construction of metastable dS vacua, such as the KKLT  \cite{Kachru:2003aw}
and the LVS \cite{Balasubramanian:2005zx} scenarios.
In light of the recent Swampland conjectures questioning the existence of controlled de Sitter vacua on general grounds~\cite{Danielsson:2018ztv, Obied:2018sgi, Garg:2018reu, Ooguri:2018wrx}, doubts concerning the correctness of the conventional 4d EFT treatment of gaugino condensation \cite{Moritz:2017xto} have received significant attention. Triggered by this, progress has been made in understanding D7-brane gaugino condensation and its interplay with dS uplifts from a 10d perspective \cite{Hamada:2018qef, Kallosh:2019oxv, Hamada:2019ack, Carta:2019rhx, Gautason:2019jwq, Bena:2019mte, Kachru:2019dvo}. While the qualitative conclusion appears to support the established 4d supergravity results (see however \cite{Gautason:2019jwq}), important details of the 10d approach have remained open. The present paper reports further progress, related specifically to the crucial quartic gaugino term. This term  receives divergent contributions from integrating out the 3-form flux and is plagued by non-localities in all available approaches. By contrast, we will be able to propose a perfectly local and covariant renormalization of this term.

Let us start by recalling the crucial pieces of the 4d ${\cal N}=1$ supergravity lagrangian as it arises in type-IIB and many other settings (see e.g.~\cite{Wess:1992cp}, App.~G):
\al{\label{4dSugra}
{\cal L}=\ldots - e^{K}\left[\left|D_a W+\frac{e^{-K/2}}{4}\left(\partial_a f\right)\lambda^2\right|^2-3|W|^2\right]+\ldots
}
We have focussed on the terms involving the superpotential $W$ and the gaugino bilinears $\lambda^2\equiv \lambda^\alpha\lambda_\alpha$. The gauge 
kinetic function and the Kahler potential are denoted by $f$ and $K$ respectively and the index $a$ labels the complex scalars. The first term in \eqref{4dSugra} is the square of a co-vector in moduli space, defined using the inverse Kahler metric: $|V_a|^2\equiv K^{a\bar b}V_a\ol{V}_{\bar{b}}$. This 4d effective Lagrangian represents the state-of-the-art for treating Kahler moduli stabilization based on gaugino condensation. Yet, somewhat surprisingly, its origin in a fundamental 10d local action of type-IIB superstrings is still obscure. Our focus will be on the derivation of the quartic gaugino term, the missing piece of the puzzle, from such a fundamental action.

In the type IIB setups of KKLT and LVS, a supergravity model of the type discussed above is engineered by wrapping stacks of D7-branes on holomorphic 4-cycles of a (warped) Calabi-Yau compactification. The superpotential is induced by 3-form fluxes which stabilize complex structure moduli and the axio-dilaton at a high energy scale. At lower energies only Kahler moduli are relevant and the superpotential is simply a constant, $W=W_0$. In case of a single Kahler modulus $T$, the index $a$ takes a unique value, the Kahler potential is $K=-3\log(T+\overline{T})$ and the gauge kinetic function is $f=T$. Jumping somewhat ahead, we note that corrections to $W$, $K$ and $f$ may in principle arise and be relevant in concrete models. Thus, while we are at the moment focussing on a 10d derivation using the familiar tree-level expressions for those quantities, the long term goal is to employ the 10d approach to control or explicitly calculate corrections.

Perturbatively, the no-scale form of $K$ implies that the scalar potential vanishes, $V(T)=0$. At the non-perturbative level, however, gaugino condensation in the IR generates a non-zero potential. Up to sub-leading quantum corrections this can be described by simply assigning a vev to the 
gaugino bilinears $\lambda\lambda \to \langle \lambda\lambda \rangle = -4a\,e^{K/2}Ae^{-aT}$ in the Lagrangian~\eqref{4dSugra}. This immediately gives the non-perturbative potential
\be\label{potential}
V(T)=e^K\Big\{\Big| D_T W_0-aAe^{-aT}\Big|^2-3|W_0|^2\Big\}\,,
\ee
which corresponds to the leading order part (in a large volume expansion) 
of the scalar potential of an ${\cal N}=1$ model with $K=-3\log(T+\overline{T})$ and $W=W_0+Ae^{-aT}$. As such, the potential \eqref{potential} and its multi-moduli generalizations (to be discussed later) are at the heart of the KKLT and LVS proposals.~\footnote{ 
We emphasize, however, that our discussion of \eqref{potential} serves mainly to motivate our investigation. The technically important equation to be matched with our 10d proposal is \eqref{4dSugra}.
}

Under appropriate circumstances, such setups have the potential to be uplifted to a dS minimum by a small source of positive energy, e.g.~an anti-D3-brane at the tip of a warped throat. Whether the right balance of ingredients can be achieved in controlled regimes of string theory has been the subject of many interesting studies recently, see e.g.~\cite{Bena:2018fqc, Blumenhagen:2019qcg, Demirtas:2019sip, Randall:2019ent, Demirtas:2020ffz, Gao:2020xqh, Bena:2020xrh, Hebecker:2020ejb, Carta:2021sms, Carta:2021lqg, Andriot:2021gwv, Honma:2021klo, Bena:2021wyr, Seo:2021kyi, DeLuca:2021pej, Bento:2021nbb}. We will not address these issues, including the particularly critical `singular bulk problem' \cite{Gao:2020xqh}, which derives from the `throat-gluing problem' of \cite{Carta:2019rhx} and was recently treated further in \cite{Carta:2021lqg, Andriot:2021gwv}. This is an issue specifically of KKLT and appears not to affect the LVS.

Our goal is simpler but at the same time more fundamental: we want to write down the type-IIB 10d action, with an 8d gauge sector localized on D7-branes, from which the 4d effective action~\eqref{4dSugra} arises upon compactification on a CY with fluxes. Such an action has long been known up to terms quadratic in fermions \cite{Camara:2004jj, Koerber:2007xk, Baumann:2009qx, Baumann:2010sx, Dymarsky:2010mf,Grana:2020hyu}. Its completion to the quartic gaugino order, a prerequisite for deriving the 4d action which stabilizes Kahler moduli, is however not properly understood.

A key problem in dealing with quartic gaugino couplings on D7-branes is the appearance of divergences. Indeed, the known coupling $\sim G_3\Psi_{D7}^2$  between bulk 3-form flux $G_3$ and localized gauginos $\Psi_{D7}$ introduces, upon integrating out the flux, a term in the 10d Lagrangian which is quartic in the 8d gaugino fields, $\sim \Psi_{D7}^4$, and becomes singular on the brane. Similar divergences were resolved for the codimension-one branes of Horava-Witten theory and related 5d toy models long ago~\cite{Horava:1995qa, Horava:1996ma, Horava:1996vs, Mirabelli:1997aj}. An action with a perfect square structure, typical of supersymmetric theories, was sufficient to cancel all divergences in such setups. The situation appears to be similar in 6d toy models with non-curved codimension-two branes \cite{Falkowski:2005zv}, although the analysis was not completed at the level of 4-fermion-terms in 4d supergravity. Curved codimension-two branes in string theory, which is our present case of interest, are much more subtle. This case has only recently been addressed in~\cite{Hamada:2018qef, Kallosh:2019oxv, Hamada:2019ack, Bena:2019mte, Kachru:2019dvo}, where finite quartic-gaugino 10d actions were proposed. Such actions lead to the right structure upon compactification to 4d (at least in the single Kahler modulus case). They can then be used to show explicitly that the generation of the potential~\eqref{potential} due to gaugino condensation can be accurately described directly in 10d language.\footnote{Such a UV description of IR effects is subtle: 
The condensing 4d gaugino is the zero-mode of an 8d field and has a fixed profile. Thus, giving a vev to gaugino bilinears yields effectively a non-local theory, which has to be handled with caution. Even if this microscopic 10d description of the physics after gaugino condensation can work only to a certain degree, it would be 
an essential tool to address issues like the interplay of non-perturbative effects and other ingredients of string compactifications (e.g. fluxes and anti-branes). The subtleties mentioned here are of course absent in the low energy 4d description. Hence the conservative reader should view our efforts merely as identifying the fundamental, local 10d action from which, by dimensional reduction, a 4d action follows in which gaugino condensation and a possible uplift can be discussed.\label{foot:locality}}

Unfortunately, the actions proposed so far are not completely satisfactory. While they successfully provide appropriate quartic gaugino couplings free of divergences, they do so at the cost of introducing non-localities in the UV action. This applies both to the perfect-square proposal of \cite{Hamada:2018qef, Hamada:2019ack}, where a certain projection operator introduces a non-locality, as well as to the analysis of \cite{Kachru:2019dvo}, where a manifestly non-local D7-brane quartic gaugino term is used. At the fundamental level, such non-localities are unacceptable: At best, these actions can be understood as approximations where effects that should be localized to D7-branes have been {\it smeared} over the whole internal space. As such, they can be useful and accurate when integrated over the internal dimensions (and hence when reproducing the 4d action~\eqref{4dSugra}), but they cannot be taken as a complete UV description of localized D7-branes.\footnote{One should distinguish these non-localities, which  must be absent in a fundamental UV action, from those mentioned in footnote~\ref{foot:locality}. The latter are the price to pay when trying to incorporate a non-perturbative IR effect in an otherwise weakly-coupled 10d UV theory, but they are harmless at the conceptual level.} While these approximate descriptions may be adequate for minimal constructions, they do not accurately capture the interactions among the various localized sources and the background fields in variants of these constructions (in some cases, such variants seem necessary, e.g. in resolving spacetime singularities \cite{Gao:2020xqh,Carta:2021lqg}).
Furthermore, such proposals are difficult to generalize to CY compactifications with multiple K\"ahler moduli.

Motivated by this situation, we construct in the present paper a fully local, finite, quartic action for gauginos localized on D7-branes and coupled to 10d bulk fields. As we show, a crucial role is played by the intersection and self-intersection-structure which is typical for D7-brane stacks. Such intersections are absent in the simpler case of codimension-one branes (e.g.~in Horava-Witten theory) but are important in generalizing the perfect-square-type action to branes with codimension $\geq 2$.

The rest of the paper is organized as follows: In Section~\ref{sec:CYaction} we isolate the divergences of the quadratic action and argue that their structure is intimately related to the (self) intersections of the relevant D7-brane stacks. With this geometrical insight, we propose a renormalization of the action: Divergent terms are naturally replaced by finite pieces, quartic in gauginos and proportional to the area of the brane self-intersections. In Section~\ref{sec:covariantaction} we give a local and 8d-covariant expression for the renormalized action. This action is the main result of this paper. In Section~\ref{sec:reduction} we explicitly show that, upon compactification on CY manifolds with arbitrary hodge numbers and D7-brane stacks, our action appropriately reduces to the well-known 4d supergravity action~\eqref{4dSugra}. We conclude in Section~\ref{sec:conclusions} and relegate technical details of spinor manipulations to the appendices.

\section{A finite local action}\label{sec:CYaction}
Our setup is type IIB string theory on a CY threefold $X$ with stacks of D7-branes that wrap a set of holomorphic 4-cycles $\Sigma_i$. Our goal is to understand the dynamics of the localized 8d gauginos $\Psi_i$ and their interactions with bulk fields, in particular the 3-form flux $G=F_3-\tau H_3$. We will make several simplifying assumptions which will help us present our results in the cleanest possible way. 

We will assume a constant axio-dilaton $\tau=C_0+i\,e^{-\phi}$ and a constant warp factor. While we are ultimately interested in situations where the profiles of these quantities are non-trivial, they do not play a decisive role in our analysis. With some extra effort such profiles can be included \cite{Giddings:2005ff,Shiu:2008ry,Douglas:2009zn} (see \cite{Marchesano:2008rg,Marchesano:2010bs,McGuirk:2012sb} for additional subtleties for brane fermions), but their presence would only complicate our expressions and obscure the key points of our discussion. Similarly, we will not consider the backreaction of the internal geometry on gaugino bilinears, which may for example be studied using Generalized Complex Geometry \cite{Koerber:2007xk, Dymarsky:2010mf,Grana:2004bg, Lust:2008zd, Bena:2019mte, Kachru:2019dvo, Grana:2020hyu}. Again, we believe that taking this into account is not necessary for our purpose which is to derive~\eqref{4dSugra}, whether gauginos condense or not, from a local 10d action. The technical reason is that we work only up to quadratic order in the operator set $\{\,\,G\,,\,(\Psi_i)^2\,\,\}$. Neither of these operators appears linearly in the action. In saying that, we do not count the gaugino kinetic term: This is a $D$-term and it will hence not develop a VEV due to the SUSY-preserving phenomenon of gaugino condensation. This gives us the right to restrict attention to field configurations where its VEV is zero. Thus, the operators from the set $\{\,\,G\,,\,(\Psi_i)^2\,\,\}$ appear in the action only quadratically and any field they source, such as the gravitational background, will be at least of quadratic order in these operators. As a result, the backreaction of the gravitational background on operators involving $G$ and $(\Psi_i)^2$ will only lead to effects of order higher than $|G|^2$, $G\,(\Psi_i)^2$ or $(\Psi_i)^4$. Expanding in this set of operators is quite natural. Indeed, the imaginary-anti-self-dual as well as  the $(0,3)$ components of the flux $G$ vanish at zeroth gaugino order by the SUSY F-term equations, as is well known in the KKLT scenario where the flux $G_{(0,3)}$ (or equivalently the superpotential $W_0$) is of order of $(\Psi_i)^2$. The $(2,1)$ components is not constrained by supersymmetry and could in principle be larger. Nevertheless, we have the right to consider a power-counting rule which assumes dilute flux and treats the operators $G$ and $(\Psi_i)^2$ as having the same order of smallness. Once we have derived the desired local $(\Psi_i)^4$ term in this fashion, it can of course be applied in situations with other parametric relations between the various fields (alternatively, we could simply consider cases where $G_{(2,1)}$ vanishes at zeroth gaugino order, but such a strong restriction is not necessary).  Finally, we will ignore gauge indices and treat gaugino fields as if their gauge group were abelian. This amounts to a restriction to situations with multiple, possibly intersecting single D7 branes. We expect that the additional step from here to the relevant case with unbroken non-abelian groups proceeds in accordance with 4d supergravity expectations.

Having said all that, we do not by any means want to imply that closing all the gaps above (from constant axio-dilaton, neglected warping to backreaction and non-abelian gauge-index structure) is uninteresting. The goal of this paper is to address a particular problem, namely renormalizing a divergent quartic gaugino term. This allows us to make these simplifications. We leave the generalizations to situations without the simplifying assumptions for future works.

Our interest is in low energies and hence in gaugino zero modes $\Psi_i=\lambda_i \otimes \chi_i$. Here $\lambda_i$ are the gauginos of the 4d SYM theory and $\chi_i$ are covariantly constant spinors on the 4-cycles $\Sigma_i$. We will be much more explicit about this decomposition later. For the moment it suffices to say that the internal spinors $\chi_i$ encode the geometry of the CY and that bilinear expressions in $\chi_i$ can be written in terms of the holomorphic 3-form $\Omega$ and the Kahler form $J$ (more specifically their restriction to the cycles $\Sigma_i$). In the present section we will obtain a 10d action to quartic gaugino order. We will write it in terms of 4d gaugino bilinears $\lambda_i^2$ and the geometric quantities $\Omega$ and $J$. Its derivation from a local, covariant 8d action depending only on $\Psi_i$ will be the subject of later sections.

\subsection{The quadratic action, its divergences, and a previous renormalization proposal}
The type-IIB action on a Calabi-Yau $X$ with D7-branes is known up to quadratic order in the gauginos \cite{Camara:2004jj, Koerber:2007xk, Baumann:2009qx, Baumann:2010sx, Dymarsky:2010mf}, the terms relevant for our analysis being 
\be\label{eq:quadratic}
\mathcal{L}_{\lambda^2}= - \frac{1}{4} \, e^{\phi} \int_X G\wedge \ast\overline{G} + \frac{1}{2}e^{\phi/2} \sum_i \left(\overline{\lambda}_i^2 \int_X G\wedge *\Omega \, \,\delta^{(0)}_i +{\text{c.c.}}\right)- \frac{i}{4}\,e^{\phi}\,\int_X G\wedge \overline{G}\,.
\ee
We work in Einstein frame and set $\kappa_{10}^2=1$. Moreover, we do not perform the integration over $\mathbbm{R}^{1,3}$ such that our expression has to be viewed as a contribution to the 4d Lagrangian. We are only interested in 3-forms with internal indices and our hodge-star is hence six-dimensional.

As already mentioned, our axio-dilaton is constant. We also set $C_0=0$, such that $G=F_3-\tau H_3= F_3-ie^{-\phi}H_3$. The delta function $\delta^{(0)}_i$ is a zero-form that localizes the integral to the surface $\Sigma_i$ wrapped by the $i$-th D7-brane. We have furthermore avoided introducing a numerical constant governing the strength of the gaugino-flux coupling by absorbing this constant into the gauginos. This affects the normalization of the gaugino kinetic term, but is not relevant for the terms that we are interested in at this point (the gaugino normalization will play a role in section~\ref{sec:check}).  The last term in \eqref{eq:quadratic} is topological and characterizes the contribution of the O3-planes and curved O7s/D7s to the energy density. It can be written as an integral over $G\wedge\ol{G}$ by D3 tadpole cancellation and the BPS condition (see e.g.~\cite{Giddings:2001yu}, App.~A.2).

Next, we complete the square and introduce the notation $|\omega|^2\equiv\int\omega\wedge \ast\,\overline{\omega}\,\,$:
\be
\mathcal{L}_{\lambda^2} = -\frac{1}{4}\, e^{\phi}\left| G - 2 \, e^{-\phi/2} \sum_i \lambda_i^2\,    \delta_i^{(0)} \,  \overline{\Omega}\,  \right|^2 \!\! +\frac{1}{4}\, e^{\phi}\, \left|2\, e^{-\phi/2}\,\sum_i \, \lambda_i^2\, \delta_i^{(0)}\,\overline{\Omega}\,\right|^2 \!\! - \frac{i}{4}\,e^{\phi}\,\int_X G\wedge \overline{G}\,.
\ee
Defining the 3-form source $j=e^{-\phi/2}\sum \lambda_i^2\, \delta_i^{(0)} \,\overline{\Omega}$, we arrive at the compact expression
\be
\mathcal{L}_{\lambda^2} = -\frac{1}{4}\,e^{\phi}\left|G-2j\right|^2+e^{\phi}\,|j|^2- \frac{i}{4}\,e^{\phi}\,\int_X G\wedge \overline{G}\,.
\ee

It is easy to see from~\eqref{eq:quadratic} how the 3-form flux is sourced by the gaugino bilinears. The equations of motion and Bianchi identity for $G$ read 
\be\label{eom}
d\ast G= 2 \, d\,\ast j =  2\,e^{-\phi/2}\sum_i \lambda_i^2\, d\left(\ast \,\overline{\Omega} \, \delta_i^{(0)}\right)\,,\qquad\qquad d\,G=0\,.
\ee
The solution is simply
\be\label{eq:solution}
G= G^{(0)}+ 2 P_e(j)=G^{(0)}+2\,e^{-\phi/2}\,\sum_i \lambda_i^2 \, P_e(\delta^{(0)}_i\,\overline{\Omega})
\ee
where $G^{(0)}$ is a harmonic form independent of $\lambda^2$ and $P_e(\omega)$ refers to the exact part in the Hodge decomposition of $\omega$. We take $G^{(0)}$ to be the quantized 3-form flux underlying our type-IIB flux compactification.

A problem arises if one attempts to use~\eqref{eq:quadratic} to reproduce the 4d effective action~\eqref{4dSugra}, including the quartic gaugino terms. Upon inserting the solution~\eqref{eq:solution} into~\eqref{eq:quadratic} one obtains a badly divergent on-shell action. Infinities arise from the term $|j|^2\sim |\lambda^2|^2(\delta^{(0)})^2$ and from the term $|G-2j|^2\sim |2P_c(j)|^2$, where $P_c(j)$ is the co-exact part of the source $j$. Here we have neglected finite, harmonic pieces of $G$ and of the Hodge decomposition of $j$. We also recall from \cite{Hamada:2018qef} that $P_c(j)$ diverges as $1/z^2$ when approaching the brane locus in the transverse plane parameterized by $z\in \mathbbm{C}$. Moreover, this divergence has an obvious fundamental origin in the impossibility to compensate the co-exact part of $2j$ by a field strength $G$ (which is closed by the Bianchi identity).

This is not too unexpected nor, in principle, too problematic, since the action~\eqref{eq:quadratic} is an expansion to first order in $\lambda^2$, while the divergences arise at quartic gaugino order $|\lambda^2|^2$. One can add quartic gaugino counterterms $\Delta{\cal L}_{|\lambda\lambda|^2}\sim |P_c(j)|^2-|j|^2$ to the Lagrangian to cancel the divergences. This was done in~\cite{Hamada:2018qef} by proposing a renormalized version of~\eqref{eq:quadratic} of the form
\be\label{eq:nonlocal}
\mathcal{L}_{\lambda^4\,,\mbox{\scriptsize \cite{Hamada:2018qef}}} =
-\frac{1}{4}\,e^{\phi}\Big|G-2(1-P_c)(j)\Big|^2- \frac{i}{4}\,e^{\phi}\,\int_X G\wedge \overline{G}\,.
\ee
This action is perfectly finite and, upon reduction to 4d, reproduces the correct parametric dependence of the 4d supergravity Lagrangian~\eqref{4dSugra}. It can be then used to study gaugino condensation from a 10d perspective. 

Unfortunately, this proposal is not completely satisfactory. The reason is that the projection operator $(1-P_c)$ which removes the co-exact piece from the gaugino source is non-local. A similar non-locality appears in \cite{Kachru:2019dvo} (cf.~App.~A.1), where the brane-localized quartic gaugino term is taken to depend on the brane-transverse volume. We believe that a completely local 10d action should exist and has to be found.

Before proceeding, let us mention that one can replace $\delta_i^{(0)}$ in the previous expressions by the two-form $\delta_i^{(2)}\in H^2(X, \mathbb{Z})$, which is Poincare-dual to the surface $\Sigma_i$. The two are related by $\delta^{(0)}=g^{m\bar{n}}\delta^{(2)}_{m\bar{n}}=i\,J \cdot \delta^{(2)}$. As an example, the source term in the quadratic action (the second term in \eqref{eq:quadratic}) can be written in several equivalent ways, e.g. 
\al{
\ol{\lambda}_i^2 \int_X G\wedge \ast \Omega\,  \delta_i^{(0)} = i\, \ol{\lambda}_i^2 \int_X \left(G\wedge \ast \Omega\right)\,  (J\cdot \delta_i^{(2)}) =  i \, \overline{\lambda}_i^2 \int J\wedge J \wedge \delta_i^{(2)} \left(G\cdot \Omega\right).
}
We could use either expression in the following analysis, but we will find it convenient to work with the zero-form $\delta_i^{(0)}$ since it provides the most straightforward way to write down the equations of motion~\eqref{eom} and the solution~\eqref{eq:solution}.
The two-form $\delta_i^{(2)}$ will become useful at the end of the calculation to renormalize the divergent on-shell action.

\subsection{Proposal for a local renormalization}

Our goal in this section is to provide an improved renormalization of~\eqref{eq:quadratic} which respects locality. It will be useful to decompose the 3-form flux into its imaginary self- and anti-self-dual parts (ISD and IASD, respectively): 
\be
G_{\pm}=\frac{1}{2} (1\mp i\,\ast)\,G	\qquad \Longrightarrow \qquad \ast \,G_{\pm}=\pm \, i\, G_{\pm}\,.
\ee
The holomorphic 3-form $\Omega$ is itself IASD, $\ast \Omega = -i\,\Omega$, such that the source $j=e^{-\phi/2}\sum \lambda_i^2\, \delta_i^{(0)} \,\overline{\Omega}$ is ISD, $\ast\,j=i\,j$. 

The quadratic Lagrangian~\eqref{eq:quadratic} can now be rewritten as
\al{\label{Eq:Lagrangian}
2e^{-\phi}\mathcal{L}_{\lambda^2}
&= - \frac{1}{2} |G|^2 +\left(\int G\wedge *\overline{j} +{\text{c.c.}}\right)- \frac{i}{2}\,\int G\wedge \overline{G}
\\
&= - \frac{1}{2} |G_+|^2 - \frac{1}{2} |G_-|^2 +\left(\int G_+\wedge *\overline{j} +{\text{c.c.}}\right) + \frac{1}{2} |G_+|^2 - \frac{1}{2} |G_-|^2
\nonumber\\
&= \left(\int G_{+}\wedge *\overline{j} +{\text{c.c.}}\right)  - |G_{-}|^2\,.
\nonumber
}
Here we have used the fact that ISD and IASD 3-forms are orthogonal w.r.t. the scalar product $(A,B)\equiv \int A\wedge \ast\overline{B}$. 

We proceed by integrating out the 3-form flux $G$, i.e.~we replace $G$ in the last line of \eqref{Eq:Lagrangian} by the corresponding solution. This solution has already been given in~\eqref{eq:solution}: $G=G^{(0)}+2\,e^{-\phi/2}\,\sum_i \lambda_i^2 \, P_e(\delta^{(0)}_i\,\overline{\Omega})$. In order to evaluate the exact part of $\delta_i^{(0)}\overline{\Omega}$, we employ the complex version of Hodge's theorem  (see e.g.~\cite{nakahara2003geometry}):
\al{
\delta_i^{(0)}\overline{\Omega}=\bar\partial \alpha_i +
 \bar\partial^\dagger (\cdots)+\omega_i=
\bar\partial \alpha_i +\omega_i=\bar\partial \alpha_i + \frac{1}{V_{i,\perp}}\overline{\Omega}\,.
\label{cht}
}
Here we used the fact that $\delta_i^{(0)}\overline{\Omega}$, being a $(0,3)$-form, is $\bar\partial$-closed and hence has no $\bar\partial^\dagger$-exact part. Moreover, $\alpha_i$ is a $(0,2)$-form and $\omega_i$ is a $\bar\partial$-harmonic $(0,3)$-form. On Kahler manifolds this is equivalent to $\omega_i$ being harmonic. The last equality in \eqref{cht} then follows because $\overline\Omega$ is the unique such form on a CY threefold. The normalization is given by 
\al{
\frac{1}{V_{i,\perp}}=\frac{V_{\Sigma_i}}{V_X}=\frac{\left(\Omega, \delta^{(0)}_i\Omega\right)}{|\Omega|^2}\,,
}
where $V_X$ and $V_{\Sigma_i}$ are the CY and the $D7_i$-brane volumes. One may think of $V_{i, \perp}$ as the ``transverse volume'' associated with $D7_i$. With this, we can extract the exact piece of $\delta_i^{(0)}\overline{\Omega}$:
\al{
P_e\left(\delta_i^{(0)}\overline{\Omega}\right)&=P_e\left(\bar\partial \alpha_i\right)=P_e\left[ \frac{1}{2}\underbrace{\left(\partial +\bar{\partial}\right)}_{\displaystyle d}\alpha_i -\frac{1}{2}\underbrace{\left(\partial -\bar{\partial}\right)}_{\displaystyle \sim \,d^\dagger\ast}\alpha_i\right]
\nonumber\\
&=\frac{1}{2}\left(\partial+\bar\partial\right)\alpha_i
=\underbrace{ \frac{1}{2} \left(\delta_i^{(0)}-\frac{1}{V_{i,\perp}}\right)\overline{\Omega}}_{\rm ISD}\,+\underbrace{\phantom{\bigg(}\!\!\!\frac{1}{2}\partial\alpha_i}_{\rm IASD}\,.\phantom{\frac{\Big(}{2}}
}
Now we can make the ISD part of $G$ completely explicit,
\al{
G_+&=G_+^{(0)}+ 2\,e^{-\phi/2}\sum_i\lambda_i^2 \left[P_e\left(\delta_i^{(0)}\overline\Omega\right)\right]_+=G_+^{(0)}+e^{-\phi/2}\sum_i\lambda_i^2\left(\delta_i^{(0)}-\frac{1}{V_{i,\perp}}\right)\overline{\Omega}\,,
}
and express its singular part in terms of the gaugino source:
\be
G_+=\left(\,G_{+}^{(0)} - e^{-\phi/2}\sum_i \frac{\lambda_i^2}{V_{i,\perp}} \, \overline{\Omega}\,\right)\,+\,j\,.\label{gpe}
\ee

This allows us to evaluate \eqref{Eq:Lagrangian} as follows:
\al{\label{fte}
2e^{-\phi}{\cal L}_{\lambda^2}&= \left(\int G_{+}\wedge *\overline{j} +{\text{c.c.}}\right)  - |G_{-}|^2 \nonumber\\
&\hspace*{-0.5cm}= - |G_+|^2+ \left(\int G_{+}\wedge *\overline{j} +{\text{c.c.}}\right)  + |G_+|^2- |G_{-}|^2 \\
&\hspace*{-0.5cm}=-\left|\,G_{+}^{(0)} - e^{-\phi/2}\sum_i \frac{\lambda_i^2}{V_{i,\perp}} \, \overline{\Omega}\,\right|^2+ |G_+^{(0)}|^2- |G_{-}^{(0)}|^2 +  e^{-\phi}\sum_{i,j} \lambda_i^2 \overline{\lambda}_j^2 \int\delta_i^{(0)} \delta_j^{(0)}  \Omega\wedge\ast\overline{\Omega}\,.\nonumber
}
Here we have used \eqref{gpe} to rewrite the first two terms in the second line. For the two final terms in the second line we have used the identity $|G_{+}|^2 - |G_{-}|^2 = |G^{(0)}_{+}|^2 - |G^{(0)}_{-}|^2$. This follows from the fact that the Chern-Simons-type term $\int G\wedge \overline{G}=|G_{+}|^2 - |G_{-}|^2$ is topological. It hence remains unchanged if the field-strength $G$ is replaced by its quantized harmonic part: $\int G\wedge \overline{G} = \int G^{(0)}\wedge \overline{G^{(0)}}$. The virtue of expressing the quadratic Lagrangian as in the last line of~\eqref{fte} is that the divergence has been cleanly isolated in its last term.

In summary, we have so far achieved the following: Starting from the Lagrangian \eqref{eq:quadratic} quadratic in the gauginos, we have integrated out the three-form flux and obtained an expression with quadratic and quartic gaugino terms. This expression~\eqref{fte}, has the right structure for reproducing the supergravity action~\eqref{4dSugra}. The only remaining problem is the divergent last term.

The rest of this paper is devoted to the renormalization of this divergence, so it is worthwhile making a few comments about its origin. The appearance of a $\delta$-square term is of course not expected in the action of a consistent effective field theory. In this case, it arises because we are treating the source (the D7-brane) as a perfectly localized object. Similar divergences are ubiquitous in field theory and appear whenever a perfectly localized source is linearly coupled to a dynamical gauge field. One example analogous to ours is that of Horava and Witten. An even more familiar case is that of the electron self-energy. Such setups are treated by standard regularization and renormalization methods, which remove any dependence on the microscopic details of the theory from the effective IR description.

It is clear, then, that the appearance of a divergence in~\eqref{fte} does not reflect problems in the original action or inappropriate manipulations of the latter. It has rather a clear physical origin in the vanishing brane thickness. As we have mentioned repeatedly, it is also not problematic at the conceptual level: Our starting point was a Lagrangian up to quadratic order in the gauginos, while the divergence arises only at quartic order. We have no a-priori knowledge of the 10d quartic gaugino action and may take it to contain a counterterm removing the divergence. In this respect, our result~\eqref{fte} already represents significant progress compared to \cite{Hamada:2018qef, Hamada:2019ack}: We have been able to achieve a formulation where the divergence is {\it completely local} (the square of a $\delta$ function), while our previous expressions contained functions diverging as one approaches the 7-branes.

The most naive expectation is that one may simply drop the divergent last term in~\eqref{fte} by appealing to the counterterm logic above. However, this turns out to be incorrect.  As we will see, the final term of~\eqref{fte} should rather be replaced by a finite expression that will be crucial for reproducing the 4d supergravity gaugino action~\eqref{4dSugra}. This is not unexpected since, whenever renormalization is required, some freedom concerning the related finite terms is unavoidably also introduced.

The term in question contains a double sum over D7 branes labelled by $\{i,j\}$. Let us first focus on the contributions with $i\neq j$. These are finite and may be evaluated in terms of the volume ${\mathcal K}_{ij}$ of the intersection locus $\Sigma_i \cap \Sigma_j$:
\al{\label{eq:regularization}
 \int \delta_i^{(0)} \delta_j^{(0)}\, \Omega\wedge*\ol{\Omega} = \int \delta_i^{(0)} \delta_j^{(0)}\, J\wedge J \wedge J = 3!\int \delta^{(2)}_i \wedge \delta^{(2)}_j \wedge J \equiv 3! \, \mathcal{K}_{ij}\,.
}
Here we made use of our convention that $\Omega\wedge*\ol{\Omega} \,/\, 3! = J\wedge J \wedge J \,/\, 3!$ is the volume form.

We would like to generalize \eqref{eq:regularization} to the problematic situation in which $i=j$. Almost all the expressions in \eqref{eq:regularization} are ill-defined in this case and are related to each other merely by formal manipulations. However, the final result $3!{\mathcal K}_{ii}$ allows for a finite interpretation: We may take it to be the volume of the self-intersection two-cycle of the four-cycle $\Sigma_i$. This two-cycle can be thought of as the locus at which two infinitesimally different representatives $\Sigma_i$ and $\Sigma_i'$ of the same homology class intersect.

Maybe more elegantly, we may define 
\be
{\cal K}_{ij}=\int [\Sigma_i]\wedge [\Sigma_j] \wedge J\qquad \quad \mbox{for both} \quad\qquad i\neq j\quad\mbox{and}\quad i=j\,.
\ee
Here $[\Sigma_i],\, [\Sigma_j]\in H^2(X, \mathbb{Z})$ are arbitrary smooth two-forms Poincare-dual to $\Sigma_i$ and $\Sigma_j$ respectively. With this definition of ${\cal K}_{ij}$, the expression \eqref{eq:regularization} for $i=j$ can be naturally interpreted as a renormalization of the case $i\neq j$. It follows by replacing the singular forms $\delta^{(2)}_i, \,\delta^{(2)}_j$ with smooth two-forms in the respective cohomology classes.  We would like to emphasize that this and the previous paragraph do {\it not} represent an evaluation of the divergent $i\!=\!j$ terms to a finite expression. The divergence is real and these two paragraphs serve only as a motivation for what the finite answer, to be obtained after renormalization, should be.

The renormalization we propose then consists in subtracting the divergent $i\!=\!j$ terms and adding finite contributions ${\cal K}_{ii}$ as defined above. With this procedure we obtain the on-shell effective Lagrangian
\al{\label{Eq:Lagrangian_Finite}
2 e^{-\phi} \mathcal{L}_{eff, \, \lambda^4}
&= - \left|G^{(0)}_{+} -  e^{-\phi/2} \sum_i {\lambda_i^2\over V_{i,\perp}} \,\overline{\Omega}\right|^2
+ |G^{(0)}_{+}|^2 - |G^{(0)}_{-}|^2
+ 3! e^{-\phi}\sum_{i,j} \lambda_i^2 \, \overline{\lambda}_j^2 \, {\cal K}_{ij}\,\,.
} 

Let us stress again that this is {\it not} the result of simply integrating out the 3-form flux from the quadratic action. Such a procedure yields a strictly {\it divergent} result, eq.~\eqref{fte}. The divergence is a consequence of treating the D7-brane sources as perfectly localized. Our proposal to replace the infinite quartic piece by the finite objects ${\cal K}_{ii}$ is a proper renormalization procedure in which a divergent counterterm needs to be added to the action, leaving behind a finite piece insensitive to the microscopic thickness of the brane.  Such renormalization is necessary in order to interpret eq.~\eqref{eq:regularization} in the case $i=j$ and then obtain~\eqref{Eq:Lagrangian_Finite} as a renormalized effective action valid at quartic gaugino order. 

The procedure described above at an intuitive level can, of course, be made completely rigorous. This requires regularization.
More precisely, all 7-brane $\delta$-functions in this paper should from the start be regularized in some standard manner, e.g. as Gaussians with a narrow width $\epsilon$. Then, as has been explained in detail, integrating out $G$ will give a $\delta$-function squared. This $\delta$-function squared will be cancelled by an appropriate counterterm in such a way that it is in the end possible to take the limit $\epsilon\to 0$, arriving at the final, renormalized  result \eqref{Eq:Lagrangian_Finite}.  Explicitly, the key step of adding the counterterm and obtaining a finite result may be formulated as
\be
 \int \!\delta_i^{(0)}(\epsilon) \,\delta_i^{(0)}(\epsilon)\, \Omega\wedge*\ol{\Omega} \,\,\rightarrow\,\,  \int \!\delta_i^{(0)}(\epsilon) \,\delta_i^{(0)}(\epsilon)\, \Omega\wedge*\ol{\Omega}
 +\Big[ 6\,{\cal K}_{ii}-
  \int \!\delta_i^{(0)}(\epsilon) \,\delta_i^{(0)}(\epsilon)\, \Omega\wedge*\ol{\Omega}
  \Big]\,.\label{eps}
\ee
Here we have suppressed all terms of the 4d lagrangian except for the divergent term with index $i$. We have also dropped the prefactor $\lambda_i^2 \, \overline{\lambda}_i^2/2$. We have made the critical term finite by introducing $\epsilon$ as explained above and added a counterm, in square brackets, which is also finite but divergent in the limit $\epsilon\to 0$. Crucially, in the complete expression on the r.h.~side of \eqref{eps} the limit $\epsilon\to 0$ may now be taken. We will leave the $\epsilon$ dependence of our expressions implicit throughout our work. 

In the rest of the paper, we will analyze this renormalization and our proposed finite action~\eqref{Eq:Lagrangian_Finite} in detail. As we will demonstrate below, it reproduces the 4d supergravity Lagrangian. But we can do more than that: We will also be able to express the critical last term in \eqref{Eq:Lagrangian_Finite} as an integral of a local 8d Lagrangian over the brane $\Sigma_i$. Explicitly, we will see that
\be
\lambda_i^2 \, \overline{\lambda}_i^2 \, {\cal K}_{ii}\sim
\lambda_i^2 \, \overline{\lambda}_i^2 \int_{\Sigma_i} F(N_i)\wedge J \sim \int_{\Sigma_i} d^4y\,\sqrt{g}\, (\ol{\Psi}_i^c[\nabla_I,\nabla_J]\Gamma^{AB}\Gamma^{IJ}\Psi_i^c) (\ol{\Psi}_i\Gamma_{AB}\Psi_i)\,.
\label{ltop}
\ee
Here in the first step, as will be explained in detail below, ${\cal K}_{ii}$ is rewritten in terms of an integral of a local geometric quantity. Concretely, $F(N_i)$ is the curvature two-form of the normal bundle $N_i$ of $\Sigma_i$. In the second step, this local geometric quantity is expressed through objects, specifically the 8d gaugino $\Psi_i$ and the Riemannian connection $\nabla$, which are suitable ingredients in a standard 8d brane action. Thus, the procedure taking us from the well-known quadratic action \eqref{eq:quadratic} to our final result, the finite quartic action \eqref{Eq:Lagrangian_Finite}, represents a conventional renormalization process: It subtracts a local divergence, the last term in \eqref{fte}, replacing it with a finite local operator, the final expression in \eqref{ltop}.

\section{The 8d local covariant action}\label{sec:covariantaction}

Our goal in this section is to demonstrate that the renormalization procedure explained above can be implemented consistently with 10d covariance and locality. The non-trivial point is to show that the last term of the  Lagrangian~\eqref{Eq:Lagrangian_Finite} with $i=j$, i.e.~$\lambda_i^2\overline{\lambda}_i^2 {\cal K}_{ii}$, can be given as an integral of a local geometric expression over the brane. Moreover, we will write this expression in terms of 8d gauginos $\Psi_i$, avoiding the 4d gauginos $\lambda_i$ which have no place in an 8d brane Lagrangian. In short, we will show that~\eqref{ltop} holds.

\subsection{The geometric formulation of ${\cal K}_{ii}$}

We want to write ${\cal K}_{ii}=\int [\Sigma_i]\wedge [\Sigma_i]\wedge J$ as an integral of a local geometric expression over the brane. Focusing on one particular brane, the index $i$ may be dropped: $\Sigma_i \to \Sigma$. We first note 
that the Poincare dual $[\Sigma]\in H^2(X, \mathbb{Z})$ of the D7-brane divisor $\Sigma$ is identical to the first Chern class of the line bundle ${\cal O}(\Sigma)$ which defines $\Sigma$ (see e.g. Proposition 4.4.13 of~\cite{huybrechts2005complex}):
\be
c_1({\cal O}(\Sigma))=[\Sigma]\,.
\ee
From this we obtain
\be
{\cal K}_{\Sigma\Sigma}\equiv \int_X [\Sigma]\wedge [\Sigma]\wedge J=\int_{\Sigma} c_1({\cal O}(\Sigma))\wedge J = \int_{\Sigma} F(N)\wedge J\,.\label{ddj}
\ee
Here $N$ denotes the normal bundle of $\Sigma$, $F(N)$ is its curvature 2-form, and we used the identification ${\cal O}(\Sigma)\big|_{\Sigma}=N$. As a matter of principle, we are now done since $F(N)$ and $J$ are both quantities which are known as soon as the local geometry near the brane is given.

For the purpose of writing our result using 8d gauginos, we first express it in terms of the covariantly constant $SO(4)$ spinor $\chi$ in the spin bundle on $\Sigma$. Since we assume that one supersymmetry is preserved, such a spinor always exists. By our assumption we have $D_m\chi=0$, such that the spinor satisfies\footnote{We follow the conventions of~\cite{Freedman:2012zz,VanProeyen:1999ni} and present the most relevant expressions in Appendix~\ref{Spinors}.}
\be
\ol{\chi}^c[D_m,D_n]\chi=0\,.\label{comm}
\ee
The covariant derivative $D_m$ is defined to include both the Riemannian spin connection and the $U(1)$ connection, associated with the line bundle $N$, by which the D7-brane gauge theory is twisted (see e.g.~\cite{Beasley:2008dc}).\footnote{We note that $\chi$ may be interpreted in a canonical way as living in the spin bundle on $X$ (pulled back to $\Sigma$). In this case the Riemannian spin connection on $\Sigma$ and the $U(1)$ connection discussed above are merged in the spin connection on $X$.} In the present section, Latin lower-case indices label the four internal directions of $\Sigma$. Since we are dealing with spinors in curved space, we find it convenient to use frame (rather than coordinate) indices throughout the paper.

Since the structure group of the spin bundle of $\Sigma$ and the $U(1)$ commute, the curvature 2-form associated with $[D_m,D_n]$ is given by the sum of the Riemannian curvature and that of the $U(1)$ connection. Thus, \eqref{comm} implies
\be\label{Rone}
\ol{\chi}^c R_{mn}\chi+\ol{\chi}^c F_{mn}\chi=0\,,
\ee
where $F_{mn}$ are the components of the curvature 2-form $F(N)$ introduced earlier. Using the covariant derivative $\nabla_m$ of the Riemannian geometry on the brane, we have 
$R_{mn}=[\nabla_m,\nabla_n]$. This allows us to write~\eqref{ddj} as
\be
{\cal K}_{\Sigma\Sigma}= -\frac{i}{4}\int_{\Sigma} d^4y\,\sqrt{g}\, \,\ol{\chi}^c \,[\nabla_m,\nabla_n]\,\chi \, J_{kl}\,\epsilon^{mnkl}\,,
\ee
where we also used the normalization $\ol\chi^c\chi=-i\chi^\dagger\chi=-i$. We may furthermore express the (pullback of the) Kahler form $J_{kl}$ in terms of the covariantly constant spinor and $SO(4)$ $\gamma$-matrices as
\be
J_{kl}=\ol{\chi}^c \,\gamma_{kl}\,\chi\,.
\ee

We hence conclude that the quartic gaugino piece $\lambda^2 \ol{\lambda}^2{\cal K}_{\Sigma\Sigma}$ of the effective Lagrangian~\eqref{ltop} is given by
\al{\label{eq:geometry}
2\, {\cal L}_{eff,\lambda^4}\,\,\supset\,\, 3!\,\lambda^2\ol{\lambda}^2\,{\cal K}_{\Sigma\Sigma}\,\,=\,- \frac{3\,i}{2} \lambda^2\ol{\lambda}^2\int_{\Sigma}  d^4y \,\sqrt{g}\,\left( \ol{\chi}^c \,[\nabla_m,\nabla_n]\,\chi \right) \left(\ol{\chi}^c \,\gamma_{kl}\,\chi\right)\epsilon^{mnkl}\,.
}
We have managed to express the 4d effective Lagranian in terms of local quantities defined on $\Sigma$, in particular the covariantly constant SO(4) spinor $\chi$. The remaining task is to write down an 8d covariant Lagrangian in terms of 8d gauginos $\Psi$ which reduces to the above when $\Psi=\lambda\otimes \chi$.

\subsection{A covariant local expression for $\lambda_i^2 \, \overline{\lambda}_i^2 \,{\mathcal K}_{ii}$}\label{sec:quartic}
Rather than working up from~\eqref{eq:geometry} and inferring the appropriate 8d covariant action, we find it simpler to present the result and show that it indeed reduces to eq.~\eqref{eq:geometry}. We hence directly propose the 8d covariant expression\footnote{Upper-case latin indices label 8d (frame) directions along $\Sigma$. Our conventions for spinors and gamma matrices are detailed in Appendix~\ref{Spinors}.}
\be
{\cal L}_8 =\frac{3\,i}{16} \int_{\Sigma}d^4y\,\sqrt{g}\,(\ol{\Psi}^c[\nabla_M,\nabla_N]\Gamma^{KL}\Gamma^{MN}\Psi^c) (\ol{\Psi}\,\Gamma_{KL}\Psi)
\label{l8}
\ee
and show that it reduces at low energies to~\eqref{eq:geometry}. This happens for the zero mode of the 8d chiral spinor, which can be written as
\be
\Psi=\lambda\otimes \chi\,,
\ee
where we take the $SO(1,3)$ and $SO(4)$ spinors $\lambda$ and $\chi$ to be both chiral. The former is a 4d quantum operator, represented by a Grassmann number, the second is a bosonic quantity encoding the profile of the zero-mode.

From the bosonic nature of $\chi$ one can see that $\ol{\chi}\chi=0$,\footnote{This follows from $\ol{\chi}\chi=(\ol{\chi}\chi)^T=\chi^T(\ol{\chi})^T=\chi^TC^T\chi=-\ol{\chi}\chi$, where $C$ is the charge conjugation matrix, which is antisymmetric in four dimensions (cf. Appendix~\ref{Spinors}).} and by chirality $\ol{\chi}\gamma_m\chi=0$. This implies that a non-zero contribution to~\eqref{l8} can arise only if both $K$ and $L$ take values along the compact dimensions. Moreover, by our no-warping assumption, the curvature term
\be\label{Rtwo}
R_{MN}=[\nabla_M,\nabla_N]
\ee
splits into two types of contributions, those with internal and those with external indices. The latter (when $M,N=\mu,\nu$) induce highly suppressed curvature corrections to the 4d effective action which we safely neglect in the following.\footnote{
To be precise, the external contribution gives schematically a piece $\Delta{\cal L}_{eff}\sim {\cal R}_{(4)}\lambda^4$. It would be interesting (independently of its smallness) to study its role in the 4d EFT, but this would require a more systematic analysis of the complete supersymmetric D7-brane action. For instance, a term of the form $\Delta {\cal L}_8\sim (\ol{\Psi}^c [\nabla_K,\nabla_L]\Psi)(\ol{\Psi}\Gamma^{KL}\Psi^c)$ gives a contribution to the effective action of the same order $\Delta {\cal L}_{eff} \sim {\cal R}_{(4)}\lambda^4$ when $K, L$ are external indices. Yet, it vanishes when $K,L$ are internal and is hence not relevant for (nor constrained by) our analysis. To say anything meaningful about the subleading curvature corrections we would need to address the presence of such $\Delta {\cal L}_8$ in the D7-brane action, and its precise effect in the 4d EFT. That analysis is beyond the scope of this work.}
Hence equation~\eqref{l8} takes the form
\be
{\cal L}_8=\frac{3\,i}{16}\,\lambda^2\ol{\lambda}^2\,\int_{\Sigma}d^4y\,\sqrt{g}\,(\ol{\chi}^c\,[\nabla_m,\nabla_n]\,\gamma^{ab}\gamma^{mn}\chi^c)(\ol{\chi}\gamma_{ab}\chi)\,,
\label{l8ll}
\ee
where we introduced the shorthand notation $\lambda^2\equiv \ol{\lambda}\lambda$ (and hence $\bar{\lambda}^2 \equiv \ol{\lambda^c}\lambda^c$).
By a slight abuse of notation, we can simply think of $\lambda$ in this expression as the 4d Weyl spinor associated to the chiral 4-component spinor $\lambda$.

In order to relate~\eqref{l8ll} to~\eqref{eq:geometry} we make use of the following Fierz rearrangement identity
\be
(\ol{\chi}^c\,[\nabla_m,\nabla_n]\,\chi)(\ol{\chi}\gamma^{mn}\chi^c)
=\frac{1}{4}\sum_{k=0}^{4}\frac{1}{k!}(\ol{\chi}^c\,[\nabla_m,\nabla_n]\,\gamma_{a_1\ldots a_k}\,\gamma^{mn}\chi^c)(\ol{\chi}\,\gamma^{a_k\ldots a_1}\,\chi)\,,
\ee
where we sum over the basis of the $SO(4)$ Clifford algebra (note that the overall sign differs from~\cite{Freedman:2012zz} since $\chi$ are commuting spinors). A non-zero contribution arises only if $\gamma^{a_1\ldots a_k}$ is a product of two gamma matrices, i.e. $k=2$. Thus
\al{
(\ol{\chi}^c\,[\nabla_m,\nabla_n]\,\chi)(\ol{\chi}\gamma^{mn}\chi^c)
&=-\frac{1}{8}(\ol{\chi}^c\,[\nabla_m,\nabla_n]\,\gamma^{ab}\gamma^{mn}\chi^c)(\ol{\chi}\,\gamma_{ab}\,\chi)
\,.
}
Applying this to \eqref{l8ll} gives
\be
{\cal L}_8=-\frac{3}{2}\,i\,\lambda^2\ol{\lambda}^2 \int_{\Sigma}d^4y\,\sqrt{g}\,(\ol{\chi}^c\,[\nabla_m,\nabla_n]\, \chi)(\ol{\chi}\,\gamma^{mn}\,\chi^c)\,.
\ee
Finally, one only needs to use the chirality of $\chi$ to write:\footnote{Notice that this is simply the statement that $J$ is self-dual in four dimensions: $J= \ast J$.}
\be
\overline{\chi}^c \gamma^{mn} \chi = \overline{\chi}^c \gamma^{mn} \gamma_{e\ast} \chi = \frac{1}{2} ( \overline{\chi}^c \gamma_{kl} \chi )\, \epsilon^{mnkl}\,,
\ee
where $\gamma_{e\ast}$ is the chirality matrix of $SO(4)$ (i.e. the ``$\gamma_5$-matrix''). With this we have shown that ${\cal L}_8$ indeed reproduces the desired expression in~\eqref{eq:geometry}
\be
{\cal L}_8= -\frac{3\,i}{4}\,\lambda^2\ol{\lambda}^2 \int_{\Sigma}d^4y\,\sqrt{g}\,(\ol{\chi}^c\,[\nabla_m,\nabla_n]\, \chi)(\ol{\chi}\,\gamma_{kl}\,\chi^c)\,\epsilon^{mnkl}.
\ee

Summarizing, we have shown in this section that the 8d covariant and local Lagrangian~\eqref{l8} yields at low energies the finite quartic gaugino term $\lambda^2\ol{\lambda}^2{\cal K}_{\Sigma\Sigma}$. That is, integrating over the internal space and keeping only the gaugino zero mode $\Psi=\lambda\otimes \chi$ (where $\chi$ is the covariantly constant spinor along $\Sigma$ and hence $\lambda$ is the 4d gaugino) one has
\al{
{\cal L}_8&=\frac{3\,i}{16} \int_{\Sigma} d^4y\,\sqrt{g}\, (\ol{\Psi}^c[\nabla_M,\nabla_N]\Gamma^{KL}\Gamma^{MN}\Psi^c) (\ol{\Psi}\,\Gamma_{KL}\Psi) \nonumber\\
&=-\frac{3\,i}{4}\,\lambda^2 \ol{\lambda}^2\,\int_{\Sigma} d^4y\, (\ol{\chi}^c\,[\nabla_m,\nabla_n]\, \chi)(\ol{\chi}\,\gamma_{kl}\,\chi^c)\,\epsilon^{mnkl}\nonumber\\
&=3\,\lambda^2\ol{\lambda}^2\,\int_{\Sigma} F(N)\wedge J = 3\,\lambda^2 \ol{\lambda}^2\,\int_X [\Sigma]\wedge [\Sigma]\wedge J \nonumber\\
&= 3\,\lambda^2\ol{\lambda}^2\,{\cal K}_{\Sigma\Sigma}\,.
}

\subsection{The complete 8d quartic gaugino action}
We have now all the ingredients necessary to write down a complete quartic gaugino action that is local, finite, and 8d covariant. As argued, this is achieved by starting with the quadratic action ${\cal L}_{\lambda^2}$ of eq.~\eqref{eq:quadratic}, removing the quartic divergent piece ${\cal L}_{div}=\frac{1}{2} |\lambda_i^2 \delta_i^{(0)}\Omega|^2$ and replacing it with the finite term ${\cal L}_8$. A covariant expression for ${\cal L}_8$ has just been given in~\eqref{l8}, and those for ${\cal L}_{\lambda^2}$ and ${\cal L}_{div}$ are presented in eqs.~\eqref{quadraticcovariant} and~\eqref{divergencecovariant} of Appendix~\ref{app:rest}. Putting them together, we can write down our proposed quartic Lagrangian
\al{\label{finaltotal}
{\cal L}_{\lambda^4}\,=\,\,\,&{\cal L}_{\lambda^2}-{\cal L}_{div}+{\cal L}_8 \nonumber\\
&\hspace*{-1.3cm}= - \frac{1}{4} \, e^{\phi} \int_X G\wedge \ast\overline{G} - \frac{i}{4}\,e^{\phi}\,\int_X G\wedge \overline{G} - \frac{1}{2}e^{\phi/2} \sum_i \left(\int_{\Sigma_i}d^4y\,\sqrt{g}\, \ol{G}_{MNz_{i}}\ol{\Psi_i} \, \Gamma^{MN}\, \Psi_i +{\text{c.c.}}\right)\nonumber\\
&\hspace*{-.7cm} + \frac{1}{2}\sum_i\,\int_{\Sigma_i} d^4y\,\sqrt{g}\,\delta_i^{(0)}\,\left(\ol{\Psi}_i^c\,\Gamma_{MN}\,\Psi_i^c\right)\left(\ol{\Psi}_i\,\Gamma^{MN}\Psi_i\,\right)\nonumber\\
&\hspace*{-.7cm}+\frac{3\,i}{16} \,\sum_i \int_{\Sigma_i} d^4y\,\sqrt{g}\, (\ol{\Psi}_i^c[\nabla_M,\nabla_N]\Gamma^{KL}\Gamma^{MN}\Psi_i^c) (\ol{\Psi}_i\,\Gamma_{KL}\Psi_i)\,,
}
where $z_i$ in the second line is the holomorphic index normal to the cycle $\Sigma_i$.\footnote{Strictly speaking, we should also label the $M,N$ indices tangential to $\Sigma_i$ with an $i$-index, i.e. $M_i,N_i$. After all, these are 8d indices defined on $\Sigma_i$. We leave this subtlety implicit in our expressions.}
We have seen that upon CY compactification, keeping only the gaugino zero modes $\Psi_i=\lambda_i\otimes \chi_i$ and solving for the $G$-flux, this results in the effective Lagrangian~\eqref{Eq:Lagrangian_Finite}, which we reproduce here for completeness:
\al{\label{Eq:Lagrangian_Finite2}
2 e^{-\phi} \mathcal{L}_{eff, \, \lambda^4}
&= - \left|G^{(0)}_{+} - e^{-\phi/2} \sum_i {\lambda_i^2\over V_{i,\perp}} \,\overline{\Omega}\right|^2
+ |G^{(0)}_{+}|^2 - |G^{(0)}_{-}|^2
+ 3!\, e^{-\phi}\sum_{i,j} \lambda_i^2 \, \overline{\lambda}_j^2 \, {\cal K}_{ij}\,\,.
}
In more detail, this happens as follows: the solution for the 3-form flux $G_+$ develops a profile localized at the $D7_i$ brane sources~\eqref{gpe}. Inserting this solution into the second line of~\eqref{finaltotal} leads to terms quartic in 4d gauginos. These include the quartic gaugino couplings that appear in the first term in~\eqref{Eq:Lagrangian_Finite2}, i.e. those proportional to~$\sim (V_{i,\perp}V_{j,\perp})^{-1}$ (for both $i\neq j$ and $i= j$). The other quartic gaugino terms that arise from the second line of~\eqref{finaltotal} after integrating out the flux should be considered separately, depending on whether they are proportional to $\sim \lambda^2_i \ol{\lambda}_j^2$ with $i\neq j$, or proportional to $\lambda^2_i \ol{\lambda}_i^2$. The former are finite and reproduce the piece $\sim{\cal K}_{ij}$ (with $i\neq j$) in the last term of~\eqref{Eq:Lagrangian_Finite2}. The latter are divergent and are exactly cancelled by the  third line of~\eqref{finaltotal}. The only remaining piece is the final line in eq.~\eqref{finaltotal}, which generates a finite quartic gaugino term corresponding to the $i=j$ component $\sim{\cal K}_{ii}$ of the last factor in~\eqref{Eq:Lagrangian_Finite2}.

Equation~\eqref{finaltotal} is the main result of our work. We have already described in detail how we came to propose such an action. Yet, one may wonder how it relates to other approaches, such as the systematic supersymmetric completion of the DBI and CS actions at higher order. In this approach, the closed string fields are typically treated as a non-dynamical background. In particular, the backreaction of gaugino bilinears on the 3-form flux is not considered. Hence the divergences with which we are concerned (in particular the third line of~\eqref{finaltotal}) are not expected to arise. On the other hand, finite four-fermion terms can arise and perhaps the last line of~\eqref{finaltotal} could be directly reproduced (indeed similar terms appear in a recent analysis of supersymmetric brane actions~\cite{Retolaza:2021gsi}). In this case, one should presumably not think of this term as of a finite part of the counterterm. Such a re-assignment, i.e. a shift of this finite part from counterterm to tree-level lagrangian, would not affect the results of the present paper. Settling these issues requires explicit knowledge of the DBI action up to fourth order in the fermions. Since this is not yet available, we have to postpone this to future work.

What remains is to show that, indeed, this reproduces the expected expression for the gaugino couplings of 4d supergravity (eq.~\eqref{4dSugra} and its generalizations to multiple Kahler moduli and multiple gauge sectors). We do this in the next section.

\section{Reduction to 4d supergravity}\label{sec:reduction}

\subsection{Notation and conventions}\label{Sec:notation}
In order to relate~\eqref{Eq:Lagrangian_Finite2} to 4d supergravity we need to express the geometric quantities involved in terms of appropriate complex moduli. The relevant analysis without gauginos can be found in~\cite{Giddings:2001yu, Grimm:2004uq}. The notation given below follows \cite{Grimm:2004uq} with slightly different normalization conventions.

Given a basis $\{\omega_\alpha\}$ of orientifold-even $(1,1)$-forms, the Kahler form and the (cohomology classes of) $2$-form delta functions associated with cycles $\Sigma_i$ are expanded as
\al{
& [J]= v^\alpha \,[\omega_\alpha]\,,
&&[\delta_i^{(2)}]= d_i^\alpha [\omega_\alpha]\,,
&&\alpha=1,\cdots , h_+^{(1,1)}\,.
}
We define
\al{
&\mathcal{K}_{\alpha\beta\g} = \int \omega_\alpha \wedge \omega_\beta \wedge \omega_\g,
&&\mathcal{K}_{\alpha\beta}=\int \omega_\alpha\wedge\omega_\beta\wedge J = \mathcal{K}_{\alpha\beta\g} v^{\g},
\\
&\mathcal{K}_{\alpha}=\int \omega_\alpha\wedge J\wedge J = \mathcal{K}_{\alpha\beta\g} v^\beta v^\g,
&&\mathcal{K} = \int J\wedge J \wedge J = \mathcal{K}_{\alpha\beta\g} v^\alpha v^\beta v^\g=3!\,V_X\,,
}
where $V_X$ is the total volume of the internal space. The $D7_i$-brane volume reads
\al{\label{Eq:D7_volume}
V_{\Sigma_i} =\frac{1}{2!} \int \delta^{(2)}_i \wedge J \wedge J= \frac{1}{2}\, d_i^\alpha \,\mathcal{K}_\alpha \, .
}
We define a `transverse volume' $V_{i,\perp}$ by
\al{
{1\over V_{i,\perp}}\equiv \frac{V_{\Sigma_i}}{V_X}={{1\over2}\int \delta_i^{(2)}  \wedge J \wedge J \over {1\over6}\int J \wedge J \wedge J}
=3\,{d_i^\alpha \mathcal{K}_\alpha \over \mathcal{K}}.
}

As for the holomorphic $(3,0)$ form, we use the normalization
\al{
\Omega\wedge *\overline{\Omega}=J\wedge J\wedge J\qquad \Longrightarrow \qquad \left|\Omega \right|^2=\int \Omega\wedge*\overline{\Omega}=\mathcal{K}.
}
Furthermore, we define a related form $\Omega'$ which does not depend on the Kahler moduli:
\al{\label{Eq:normalization}
\Omega' \equiv {\Omega \over \sqrt{\mathcal{K}}} \qquad \Longrightarrow \qquad
\left|\Omega' \right|^2 = 1\,.
}

\subsection{The 4d supergravity Lagrangian}\label{sec:4dsugra}

To simplify our analysis we focus only on gauginos, Kahler moduli and the axio-dilaton. This means in particular that we set the complex structure moduli to their vacuum values, which corresponds to choosing the harmonic background flux $G^{(0)}$ to have only $(0,3)$ and $(3,0)$ components. Extending the analysis to complex structure moduli and hence more general fluxes with $(1,2)$ and $(2,1)$ components is standard (see e.g.~\cite{Giddings:2001yu}) and proceeds in analogy to the treatment of the axio-dilaton. We furthermore set all axionic components to zero (for details see~\cite{Grimm:2004uq}). Our restricted analysis is general enough to validate our 10d proposal. 

In type II orientifold compactifications the Kahler potential for Kahler moduli and the axio-dilaton takes the form
\al{
K=-2\log\left[{\cal{K}}(T_\alpha,\ol{T}_{\bar\alpha})\right]-\log\left[-i(\tau-\ol{\tau})\right]\,.\label{kk0}
}
We have chosen conventions in which no extra additive constant appears in \eqref{kk0}. The real parts of the holomorphic Kahler coordinates $T_\alpha$ are proportional to four-volumes: Re$\,T_\alpha=\frac{3}{4}{\cal K}_\alpha=\frac{3}{4}{\cal K}_{\alpha\beta\gamma}v^\beta v^\gamma$. From this it follows that $K_{T_\alpha}=-2v^\alpha/{\cal K}$ and $K_\tau=ie^{\phi}/2$. The Kahler metrics and their inverses are~\cite{Grimm:2004uq}
\al{
&K_{T_\alpha\ol{T}_{\beta}}=\frac{1}{{\cal K}^2}\left(-\frac{2}{3}\,{\cal K}\,{\cal K}^{\alpha\beta}+2 \,v^\alpha v^\beta \right)\,,\qquad 
&&K^{T_\alpha\ol{T}_{\beta}}=-\frac{3}{2}\left({\cal K}\,{\cal K}_{\alpha\beta}-\frac{3}{2}\,{\cal K}_\alpha\,{\cal K}_{\beta}\right)\,, \nn
&K_{\tau\ol{\tau}}=\frac{-1}{(\tau-\ol{\tau})^2}=\frac{1}{4} e^{2\phi}\,, &&K^{\tau\ol\tau}=4\,e^{-2\phi}\,.
}
The superpotential (before gaugino condensation) and gauge kinetic functions are given by\footnote{ The superpotential could have been normalized differently at the expense of introducing an additive constant in \eqref{kk0}. Similarly, the normalization of the  gauge kinetic function can be changed together with that of the gauginos. As explained after \eqref{eq:quadratic}, we have so far normalized the gauginos such as to avoid any numerical constant in the gaugino-flux coupling.
The gaugino normalization in the present subsection is different. This will be fixed momentarily.}
\al{
&W(\tau)=\int G\wedge \Omega'\,,	&&f_i=d_i^\alpha\,T_\alpha\,,
}
where $\Omega'$ (rather than $\Omega$) appears since the superpotential is independent of the Kahler moduli. One then has
\al{
D_{T_\alpha}W&=K_{T_\alpha}W\,,		&& D_\tau W = \frac{-1}{\tau-\ol{\tau}}\int \ol{G}\wedge \Omega'=\frac{i\,e^{\phi}}{2}\int \ol{G}\wedge \Omega'\,.
}
The 4d supergravity Lagrangian
corresponding to this model reads (see e.g.~\cite{Wess:1992cp}, App.~G)
\al{\label{4dcomplete}
{\cal L}_{\text{(sugra)}}=-e^K \paren{ \left|{e^{-K/2}\over4} (\p_{T_\alpha} f_i)\lambda_i^2 + D_{T_\alpha} W \right|^2
+ \left|D_\tau W \right|^2
-3|W|^2}+\,\ldots\,,
}
where sums over $i$ are henceforth left implicit and, as mentioned before, the square of a co-vector in moduli space relies on the inverse Kahler metric: $|V_{\Phi_a}|^2\equiv K^{\Phi_a\ol{\Phi}_b}V_{\Phi_a}\ol{V}_{\ol{\Phi}_b}$. We are now ready to confirm that our proposal~\eqref{Eq:Lagrangian_Finite2} takes precisely this form.

\subsection{Check of the proposal}\label{sec:check}
As mentioned, we are only keeping track of the dependence on Kahler moduli and the axio-dilaton. Hence we only consider non-zero $(0,3)$ and $(3,0)$ components of the harmonic background flux $G^{(0)}$:
\al{
&G^{(0)}= G_{(0,3)}^{(0)}\overline{\Omega}' + G_{(3,0)}^{(0)}\Omega'\,.
}
Notice that we use $\Omega'$ rather than $\Omega$ since the flux is independent of Kahler moduli. $G_{(0,3)}^{(0)}$ and $G_{(3,0)}^{(0)}$ are then just functions of $\tau$ subject to flux quantization conditions. With our normalization~\eqref{Eq:normalization} we can write
\al{
&G_{(0,3)}^{(0)}=\int G^{(0)} \wedge \ast\Omega'=-i\,W \quad &&\Rightarrow \quad |G_{(0,3)}^{(0)}|^2=|W|^2
\\
&G_{(3,0)}^{(0)}=\int G^{(0)}\wedge\ast \overline{\Omega}'=-2\,e^{-\phi}\,D_{\bar{\tau}} \ol{W} \quad &&\Rightarrow \quad |G_{(3,0)}^{(0)}|^2= K^{\tau\ol{\tau}}D_\tau W\,D_{\ol{\tau}}\ol{W}=|D_\tau W|^2\,. \nonumber
}
There is only one more subtlety that we need to take into account before comparing our proposal with the standard 4d supergravity Lagrangian. In previous sections we have worked in 10d Einstein frame and with gauginos $\Psi=\lambda\otimes \chi$ being   8d canonically normalized (up to a numerical constant mentioned after equation~\eqref{eq:quadratic}). We need to bring this to the appropriate 4d normalization. We go to 4d Einstein frame by rescaling $g_{\mu\nu}\to {\cal K}^{-1} g_{\mu\nu}$ and  normalize the 4d gauginos by $\lambda_i^2\to \sqrt{2}\,c\,{\cal K}^{3/2}\lambda_i^2$. 

After these rescalings, we can massage the action~\eqref{Eq:Lagrangian_Finite2}, using heavily the expressions presented in the previous two subsections, as follows:
\begin{eqnarray}
 &&\hspace{-.3cm}\mathcal{L}_{eff, \, \lambda^4}
= \frac{e^\phi  }{2\,{{\cal K}^{2}}}\Big[- \left|G^{(0)}_{+} -  \sqrt{2}\,c \,\K^{3/2} e ^{-\phi/2} {\lambda_i^2\over V_{i,\perp}} \,\overline{\Omega}\right|^2
+ |G^{(0)}_{+}|^2 - |G^{(0)}_{-}|^2
\nonumber\\
&& \hspace{10cm}+ 3!\,2\,|c|^2\,\K^3\,e^{-\phi}  \lambda_i^2 \, \overline{\lambda}_j^2 \, {\cal K}_{ij}\Big]
\nonumber\\
&&\hspace{-.5cm}= e^{K}\left[- \left|G^{(0)}_{+} -  3\, c\, e^{-K/2}\, \K_\alpha\, \ol{\Omega}'  d_i^\alpha \,\lambda_i^2 \,\right|^2
+ |G^{(0)}_{+}|^2 - |G^{(0)}_{-}|^2
+3!\,|c|^2\,e^{-K}\K \,\K_{\alpha\beta} d_i^\alpha d_j^{\beta} \lambda_i^2 \, \overline{\lambda}_j^2\,\right]\nonumber\\
&&\hspace{-.5cm}=e^{K}\left[3\,\ol{c}\,e^{-K/2}\K_\alpha\left(\int G_+^{(0)}\wedge \ast \Omega'\right)d_i^\alpha \ol{\lambda}_i^2+(c.c.) - |G^{(0)}_{-}|^2\right.\nonumber\\
&& \hspace{4cm}-\left.9\,|c|^2\,e^{-K}\K_\alpha\K_\beta d_i^\alpha d_j^{\beta} \lambda_i^2 \, \overline{\lambda}_j^2
+3!\,|c|^2\,e^{-K}\K \,\K_{\alpha\beta} d_i^\alpha d_j^{\beta} \lambda_i^2 \, \overline{\lambda}_j^2\,
\right]\nonumber\\
&&\hspace{-.5cm}=e^{K}\Bigg[-3i\,\ol{c}\,e^{-K/2}\K_\alpha \,W\left(\partial_{{\ol T}_\alpha}\ol{f}_i\right)\ol{\lambda}_i^2+(c.c.) - K^{\tau\ol\tau}(D_\tau W)(D_{\bar{\tau}}\ol{W})\nonumber\\
&& \qquad\qquad+
3!\,|c|^2\,e^{-K}\underbrace{\left(\K\K_{\alpha\beta}-\frac{3}{2}\K_\alpha\K_\beta\right)}_{\sim K^{T_\alpha\ol{T}_\beta}}  \left(\partial_{T_\alpha}f_i\right) \left(\partial_{{\ol T}_\beta}\ol{f}_j\right) \lambda_i^2 \overline{\lambda}_j^2\,
\Bigg]\,.
\end{eqnarray}
We recognize in the last line the appearance of $K^{T_\alpha\ol{T}_\beta}$. Using this metric and its inverse, we proceed to complete squares
\begin{eqnarray}
\hspace{-.2cm}  \mathcal{L}_{eff, \, \lambda^4}\hspace{-.5cm}
 &&=e^{K}\left[-3i\,\ol{c}\,e^{-K/2}\K_\alpha \,W\left(\partial_{{\ol T}_\alpha}\ol{f}_i\right)\ol{\lambda}_i^2+(c.c.) - |D_\tau W|^2\right.\nonumber\\
&& \hspace{6cm}-\left.
4\,|c|^2\,e^{-K}\,K^{T_\alpha\ol{T}_\beta} \left(\partial_{T_\alpha}f_i\right)\left(\partial_{{\ol T}_\beta}\ol{f}_j\right) \lambda_i^2\overline{\lambda}_j^2\,
\right]\nonumber\\
&&=e^{K}\left[-K^{T_\alpha\ol{T}_\beta}\left(2\,c \,e^{-K/2} \left(\partial_{ T_\alpha} f_i\right)\lambda_i^2 + \frac{3i}{2}W K_{T_\alpha\ol{T}_\rho}\K_\rho\right)\times\right.
\\
&&
\hspace{.5cm}
\left. \left(2\,\ol{c} \,e^{-K/2} \left(\partial_{\ol{T}_\beta} \ol{f}_j\right)\ol{\lambda}_j^2 - \frac{3i}{2}\ol{W} K_{T_\gamma\ol{T}_\beta}\K_\gamma\right) + \frac{9}{4} K_{T_\alpha\ol{T}_\beta}\K_\alpha\K_\beta |W|^2- |D_\tau W|^2
\right]\,.
\nonumber
\end{eqnarray}
It follows from the previous subsection that $K_{T_\alpha\ol{T}_\beta}\K_\beta=-\frac{2}{3}K_{T_\alpha}$ and $K_{T_\alpha\ol{T}_\beta}\K_\alpha\K_\beta=\frac{4}{3}$. With this we can further simplify the effective Lagrangian to
\al{
 \mathcal{L}_{eff, \, \lambda^4}
&=e^{K}\Big[\!-\!K^{T_\alpha\ol{T}_\beta}\Big(2c \,e^{-K/2} \left(\partial_{ T_\alpha} f_i\right)\lambda_i^2 - i K_{T_\alpha}W\Big)
\Big(2\ol{c} \,e^{-K/2} \left(\partial_{\ol{T}_\beta} \ol{f}_j\right)\ol{\lambda}_j^2 + i K_{\ol{T}_\beta}\ol{W}\Big)\nonumber\\
& \hspace{6cm}+
3 |W|^2- |D_\tau W|^2\Big]\,.
}
By taking $c=-i/8$ we see that this has precisely the form of the 4d supergravity Lagrangian~\eqref{4dcomplete}:
\al{
{\cal L}_{eff,\,\lambda^4}=-e^K \paren{ \left|{e^{-K/2}\over4} (\p_{T_\alpha} f_i)\lambda_i^2 + D_{T_\alpha} W \right|^2
+ \left|D_\tau W \right|^2
-3|W|^2}={\cal L}_{\text{(sugra)}}\,.
}
This is what we wanted to prove.

\section{Conclusions}\label{sec:conclusions}

In this work we have completed the known D7-brane action for gauginos and their coupling to 3-form flux at the quartic order in fermions. The new 4-fermion  term involves a piece, which is necessary for subtracting the divergences that arise from integrating out the flux, as well as a finite part related to the self-intersection structure of the brane.

We have arrived at this result in the following manner: First, we integrate out the complexified 3-form field strength in the presence of the familiar brane-localized gaugino-bilinear source. This results in a divergence which can be renormalized and brought into a form involving 4 gauginos and the volume of the brane-(self-)intersection locus. Alternatively, one may characterize this step by saying that we subtract a local divergence and add a finite 4-gaugino term. We show explicitly that the latter can be derived from an integral over a perfectly local, 8d-covariant brane-localized operator. 
Finally, we demonstrate that upon compactification on generic CY orientifolds the complete action appropriately reduces to the well-known action of 4d supergravity. This crucially relies on our new 4-fermion contribution.

Our approach is rather hands-on and focuses only on the terms of the action relevant for moduli stabilization (and hence for constructions of dS vacua as in KKLT and LVS). In particular, obtaining the 4d lagrangian represents our main consistency check. It would be interesting to supplement our analysis by an approach which is entirely 10-dimensional, such as the 10d Noether procedure\cite{Horava:1996ma}, or by the construction of $\kappa$-symmetric D7-brane action at the required order in fermions (see e.g.~\cite{Grana:2020hyu} for recent progress).  Very recently, \cite{Retolaza:2021gsi} provided a method to systematically expand the DBI action which is in principle suitable to extract quartic gaugino terms on D7-branes. It would be extremely interesting to understand the relation to our results, in particular the cancellation of divergences in that approach. 

Let us take stock of what we have found. It has long been known that supersymmetry demands a quartic gaugino term to complete the supergravity action to a perfect square form \cite{Dine:1985rz}. This observation carries over to gauginos on codimension one branes without any new wrinkles \cite{Horava:1995qa,Horava:1996ma,Horava:1996vs, Mirabelli:1997aj}: the perfect square structure is sufficient to cancel the naive divergences associated with the gauginos, leading to a local, finite action. We found in this work that localized sources (e.g., gaugino condensates) on codimension two branes introduce further subtleties. While the perfect square structure is still crucial, renormalizing the localized divergences requires a more careful treatment. We showed that the renormalization of the divergent D7-brane action has a geometric interpretation in terms of the self-intersection of the 4-cycles on which the D7-branes wrap. This explains why this subtlety does not arise in the codimension one case. We expect similar renormalizations would be needed for localized sources on branes with codimension $\geq 2$ which are ubiquitous in string theory. Our findings thus provide a useful first step for such future investigations.

\section*{Acknowledgments}

We would like to thank M.~Kim, V.~Van Hemelryck and T.~Van Riet for useful discussions. The work of YH is supported by JSPS Overseas Research Fellowships. The work of GS is supported in part by the DOE grant DE-SC0017647 and the Kellett Award of the University of Wisconsin. The work of AH was supported by the Deutsche Forschungsgemeinschaft (DFG, German Research Foundation) under Germany's Excellence Strategy EXC 2181/1 - 390900948 (the Heidelberg STRUCTURES Excellence Cluster). The work of PS is supported by IBS under the project code, IBS-R018-D1. He acknowledges funding from the European Union’s Horizon 2020 research and innovation programme under the Marie Sklodowska-Curie grant agreement No 690575 (RISE InvisiblesPlus) and thanks the Kavli IPMU for hospitality in early stages of this work. We would like to thank for the hospitality of the Simons Center for Geometry and Physics, where this work was initiated during the 2018 Simons Summer workshop, and of the IPhT Saclay and the KITP Santa Barbara, where further progress was made during the workshop `de Sitter Constructions in String Theory' in 2019 and the program `The String Swampland' in 2020, respectively.

\appendix

\section{Spinors in $SO(1,7)\to SO(1,3)\otimes SO(4)$}\label{Spinors}
We present in this section our conventions and several useful relations for spinors and their decomposition. We follow mostly~\cite{Freedman:2012zz,VanProeyen:1999ni} which we slightly adjust when needed (e.g. when dealing with commuting spinors).

We are interested in SO(1,7) spinors in the eight dimensions corresponding to the whole worldvolume of D7-branes (labelled by upper-case latin indices $I,J,\ldots=0,\ldots,7$); $SO(1,3)$ spinors in the four non-compact dimensions (labelled by lower-case greek indices $\mu,\nu,\ldots=0,\ldots,3$); and $SO(4)$ spinors in the four internal directions tangential to the cycle $\Sigma$ wrapped by the D7-branes (labelled by lower-case latin indices $m,n,\ldots=4,\ldots,7$). We use upper-case symbols for eight dimensional quantities, and lower-case symbols for four dimensional ones (both for $SO(1,3)$ and $SO(4)$). We use a label `$e$' (for Euclidean) when we need to distinguish $SO(4)$ ingredients from their $SO(1,3)$ analogs.

In $d$ dimensions (with $d$ even), chirality matrices (the analogs of $\gamma_5$ in 4d) are given by
\al{
&\text{Lorentzian:} \qquad \Gamma_{\ast}=(-i)^{\frac{d}{2}-1}\,\Gamma^0\ldots \Gamma^{d-1}\,;\nonumber\\
&\text{Euclidean:}   \qquad \Gamma_{e\ast}=(-i)^{\frac{d}{2}}\,\Gamma^1\ldots \Gamma^{d}\,.
}
Conjugate spinors are defined as:
\al{
\overline{\Psi}=\Psi^T C\,, \qquad \Psi^c=B^{-1}\Psi^{\ast}\,,
}
where $C$ is the charge conjugation matrix and the definition of $B$ depends on the spacetime signature:
\al{
\text{Lorentzian:}  \qquad B\equiv it_0\, C \Gamma^0 \,;\qquad \qquad \text{Euclidean:}   \qquad B_e\equiv it_1 C_e \,.
}
Here $t_0=\pm 1$ and $t_1=\pm 1$ are sign factors that determine the symmetry properties of the $\Gamma$-matrices:
\al{\label{t-signs}
 (C\Gamma_{(n)})^T=-t_{(n)}(C\Gamma_{(n)})\,,
}
where $\Gamma_{(n)}$ is the antisymmetric $\Gamma$-matrix with $n$-indices. The signs $t_{(n)}$ depend only on the dimensionality of space-time and satisfy $t_{(n+2)}=-t_{(n)}$. In particular, we have $(t_0,t_1)=(-1,-1)$ in eight dimensions; and $(t_0,t_1)=(1,-1)$ in four dimensions.

Given sets $\{\gamma^\mu\}$ and $\{\gamma^m\}$ of gamma matrices of $SO(1,3)$ and $SO(4)$, we can construct those of $SO(1,7)$ as
\al{
\Gamma^\mu&=\gamma^\mu \otimes \gamma_{e\ast} \, \qquad \mu=0,\ldots 3\,,\nonumber\\ 
\Gamma^m&= \I \otimes \gamma^m ~~~ \qquad m=4,\ldots 7\,.
}
From these, one can read off the relations
\al{
\Gamma_{\ast}&=i\,\Gamma^0\ldots \Gamma^7 = \gamma_{\ast}\otimes \gamma_{e\ast}\,,\nonumber\\
C&=\left(c\otimes c_{e}\right)\,\Gamma_{\ast} = c\,\gamma_{\ast} \otimes c_{e} \gamma_{e\ast}\,,\nonumber\\
B&=-i C \Gamma^0 = i\,b\,\gamma_{\ast}\otimes b_{e}\,.
}

We take 8d gauginos $\Psi$ to be left-handed ($\Gamma_\ast \Psi=\Psi$). Under dimensional reduction, they decompose as
\al{
\Psi=\lambda\otimes \chi\,,
}
where $\lambda$ is the 4d gaugino, and $\chi$ is a covariantly constant (commuting) $SO(4)$ spinor on the cycle $\Sigma$ wrapped by the D7-brane stack. The latter descends from the covariantly constant $SO(6)$ spinor of the 6d CY geometry. If the brane is holomorphically embedded, it easy to see that this 4d spinor is chiral~\cite{Berkooz:1996km,Beasley:2008dc,Donagi:2008ca}. We take both $\lambda$ and $\chi$ to be left-handed. In turn, the internal spinor $\chi$ encodes the geometry of $\Sigma$ through the relations
\al{\label{eq:J&Omega}
J_{mn}=\ol{\chi}^c\gamma_{mn}\chi\,, \qquad 
\Omega_{mnz}=\overline{\chi}^c\gamma_{mn}\chi^c\,.
}
where $J$ and $\Omega$ are the Kahler form and the holomorphic $(3,0)$-form of the CY, pulled back to $\Sigma$. The index $z$ labels the holomorphic direction transverse to $\Sigma$ on the CY. These are nothing but the standard relations in CY geometry between $J$ and $\Omega$ and the covariantly constant $SO(6)$ spinors (see e.g.~\cite{Strominger:1986uh}) pulled back to the holomorphic four-cycle $\Sigma$. Notice that $\Omega_{mnz}$ transforms as a co-vector under $SO(2)$ rotations in the $z$-plane. Covariance of~\eqref{eq:J&Omega} is maintained, despite the absence of manifest $z$ indices in the spinorial expressions, because the spinor $\chi$ transforms with a phase under $SO(2)$.

 Manipulations of bosonic spinors are not typically discussed in the literature, so let us collect a few useful results here. We focus on spinors of $SO(4)$. These results can be derived from the previous expressions, but it will be useful to present them explicitly to make the derivations in the rest of the paper more transparent.

One first thing to notice is that charge conjugation squares to $-1$, i.e. $(\psi^c)^c=-\psi$. This is the reason why there are no Majorana spinors in 4d Euclidean space. This is important when taking the  complex conjugation of spinor bilinears, for which the following formula is also useful
\al{
\left(\ol{\psi}_1 M \psi_2\right)^*=-\,\ol{\psi}_1^c \left(B^{-1}M^*B\right)\psi_2^c\equiv -\,\ol{\psi}_1^c M^c\psi_2^c\,,
}
for any $SO(4)$ bosonic spinors $\psi_{1,2}$ and an arbitrary matrix $M$. In particular, one can see that $\gamma_{mn}^c=\gamma_{mn}$. To keep the notation simple, we take the indices $m,n$ to be real. Similarly, the Majorana flip relation reads
\al{
\ol{\psi}_1\, \gamma_{(n)} \psi_2=-t_{(n)} \ol{\psi}_2\, \gamma_{(n)} \psi_1\,,
}
where $t_{(n)}$ are the sign factors introduced in~\eqref{t-signs}. Notice that $\ol\psi \psi=0$ follows immediately.

We can further use these relations to derive properties of the geometric quantities $J$ and $\Omega$ given in~\eqref{eq:J&Omega}. It is now easy to check that $J$ is real and satisfies
\al{
J_{mn}=\ol{\chi}^c\gamma_{mn}\chi=\ol{\chi}\gamma_{mn}\chi^c=(J_{mn})^*\,.
}
On the other hand the complex conjugate of $\Omega$ is given by
\al{\label{spinorsubtleties}
\left(\Omega_{mnz}\right)^*=\ol{\Omega}_{mn\bar{z}}=-\ol{\chi}\gamma_{mn}\chi
}
These type of expressions will be useful in the manipulations of  spinors  throughout this paper.

\section{The rest of the covariant Lagrangian}\label{app:rest}
In section~\ref{sec:quartic} we have described how to obtain the term $\lambda_i^2\ol{\lambda}_i^2 {\cal K}_{ii}$ of the effective action~\eqref{Eq:Lagrangian_Finite} from a local and covariant Lagrangian ${\cal L}_8$, eq.~\eqref{l8}, given in terms of 8d gauginos $\Psi$. As we have explained, we want to interpret this term as the renormalized version of the divergent coupling $|\lambda_i^2 \delta_i^{(0)}\Omega|^2$ implicitly contained in~\eqref{eq:quadratic} (see eq.~\eqref{fte}). In this section we write the rest of the terms in~\eqref{eq:quadratic}, including this divergent piece, in an 8d covariant manner as well.

The bulk $G$-flux kinetic and Chern-Simons terms in~\eqref{eq:quadratic} are already written in a 10d covariant manner and need no further discussion. Hence we focus on the flux-gaugino coupling first obtained in~\cite{Camara:2004jj}
\al{
\lambda_i^2 \int_X \ol{G}\wedge *\ol{\Omega} \, \,\delta^{(0)}_i +{\text{c.c.}}
=\lambda_i^2 \int_{\Sigma_i}d^4y\, \sqrt{g}\,  \ol{G} \cdot\ol{\Omega} +{\text{c.c.}}
}
Using the expressions in Appendix~\ref{Spinors} it is easy to write this coupling in terms of the (zero mode of the) 8d gaugino $\Psi_i=\lambda_i\otimes\chi_i$:
\al{\label{sign}
\lambda_i^2\,\ol{G}\cdot\ol{\Omega}&=(\ol{\lambda}\lambda)_i\, \ol{G}_{mnz_i}\,\ol{\Omega}^{mnz_i}= - \ol{G}_{mnz_i}\, (\ol{\lambda}\lambda)_i\, (\overline{\chi}_i\gamma^{mn}\chi_i) \nonumber\\
&= - \ol{G}_{mnz_i}\, (\ol{\lambda}_i\otimes \ol{\chi}_i)\,(\I\otimes \gamma^{mn})\,(\lambda_i\otimes \chi_i)=- \ol{G}_{MNz_i}\ol{\Psi}_i \, \Gamma^{MN}\, \Psi_i\,,
}
where, as before, $z_i$ labels the holomorphic direction normal to $\Sigma_i$.
With this, we conclude the rewriting of the quadratic Lagrangian~\eqref{eq:quadratic} in an 8d covariant manner:
\al{\label{quadraticcovariant}
\mathcal{L}_{\lambda^2}&= - \frac{1}{4} \, e^{\phi} \int_X G\wedge \ast\overline{G} - \frac{i}{4}\,e^{\phi}\,\int_X G\wedge \overline{G} + \frac{1}{2}e^{\phi/2} \sum_i \left(\overline{\lambda}_i^2 \int_X G\wedge *\Omega \, \,\delta^{(0)}_i +{\text{c.c.}}\right)\\
&\hspace*{-.9cm}= - \frac{1}{4} \, e^{\phi} \int_X G\wedge \ast\overline{G} - \frac{i}{4}\,e^{\phi}\,\int_X G\wedge \overline{G} - \frac{1}{2}e^{\phi/2} \sum_i \left(\int_{\Sigma_i}d^4y\,\sqrt{g}\, \ol{G}_{MNz_i}\ol{\Psi}_i \, \Gamma^{MN}\, \Psi_i +{\text{c.c.}}\!\right).\nonumber
}
Note that the last term has full 10d/8d covariance in spite of the explicit appearance of the brane-transverse index $z_i$. The reason is that $\Psi_i$ transforms by a phase under SO(2) rotations in the brane-transverse plane (consistent with the twisting of the 8d gauge theory described in~\cite{Beasley:2008dc}), as mentioned after~\eqref{eq:J&Omega}.

The same manipulations can be used to write a covariant expression for the divergent piece $|\lambda_i^2 \delta_i^{(0)}\Omega|^2$ that we want to subtract from~\eqref{quadraticcovariant}:
\al{\label{divergencecovariant}
{\cal L}_{div}&=\frac{1}{2}\sum_i\lambda_i^2\ol{\lambda}_i^2\,\int_X\delta_i^{(0)}\delta_i^{(0)}\,\Omega\wedge\ast\ol{\Omega}\nonumber\\
&=\frac{1}{2}\sum_i\,\int_{\Sigma_i} d^4y\,\sqrt{g}\,\delta_i^{(0)}\,\left(\ol{\lambda}_i^2\,\Omega_{mnz_i}\right)\,\left(\lambda_i^2\,\ol{\Omega}^{mnz_i}\right)\nonumber\\
&=-\frac{1}{2}\sum_i\,\int_{\Sigma_i} d^4y\,\sqrt{g}\,\delta_i^{(0)}\,\left(\ol{\Psi}_i^c\,\Gamma_{MN}\,\Psi_i^c\right)\left(\ol{\Psi}_i\,\Gamma^{MN}\Psi_i\,\right)\,.
}
Notice that these expressions vanish when the indices $M,N=\mu,\nu$ take values on the external directions.
  
Equations~\eqref{quadraticcovariant} and~\eqref{divergencecovariant}, together with~\eqref{l8}, are all the pieces we need to write a finite, local and 8d covariant Lagrangian, quartic in gauginos, which reproduces the 4d supergravity action~\eqref{4dSugra} upon compactification on a CY.

\bibliography{Bibliography}\bibliographystyle{utphys}

\end{document}